\documentclass[12pt,preprint]{aastex}

\newcommand{\simless}{\mathbin{\lower 3pt\hbox
      {$\rlap{\raise 5pt\hbox{$\char'074$}}\mathchar"7218$}}} 
\newcommand{\simgreat}{\mathbin{\lower 3pt\hbox
     {$\rlap{\raise 5pt\hbox{$\char'076$}}\mathchar"7218$}}} 

\slugcomment{\today}

\shorttitle{Interferometric observations of the young brown dwarf 2M0444}
\shortauthors{Ricci et al.}

\begin{document}


\title{CARMA interferometric observations of 2MASS J044427+2512: the first spatially resolved observations of thermal emission of a brown dwarf disk}


\author{L. Ricci, A. Isella, J. M. Carpenter}
\affil{Department of Astronomy, California Institute of Technology, MC 249-17, Pasadena, CA 91125, USA}

\and

\author{L. Testi}
\affil{European Southern Observatory, Karl-Schwarzschild-Strasse 2,
D-85748 Garching bei M$\ddot{u}$nchen, Germany}


\email{lricci@astro.caltech.edu}


\begin{abstract}

We present CARMA 1.3 mm continuum data of the disk surrounding the young brown dwarf 2MASS J044427+2512 in the Taurus molecular cloud. The high angular resolution of the CARMA observations (0.16$''$) allows us to spatially resolve for the first time the thermal emission from dust around a brown dwarf.
We analyze the interferometric visibilities and constrain the disk outer radius adopting disk models with power-law radial profiles of the dust surface density. In the case of a power-law index $\leq 1$, we obtain a disk radius in the range of about 15 - 30 AU, while larger disks are inferred for steeper radial profiles.
By combining this information on the disk spatial extent with the sub-mm spectral index of this source we find conclusive evidence for mm-sized grains, or larger, in this brown dwarf disk. We discuss the implications of our results on the models of dust evolution in proto-planetary disks and brown dwarf formation. 
 
\end{abstract}



\section{Introduction}
\label{sec:intro}


The investigation of disks around brown dwarfs (BDs) allows to test the physics of disk evolution and planet formation in physical environments which are very different from those characteristics of disks around young solar-like stars~\citep[][]{Luhman:2007a,Apai:2008}.  
In the case of pre-main sequence (PMS) stars, disks are formed as a consequence of the conservation of angular momentum during the gravitational collapse of the protostellar envelope. 
Most of the mass forming a newborn star/BD comes from accretion through a circumstellar disk
and characterizing the structure of accretion disks around young BDs has the potential to constrain the mechanisms of their formation \citep[see][]{Luhman:2012}.

Disk masses have been measured for a couple of tens of young BDs through far-IR and sub-millimeter photometry under the assumption of optically thin emission from dust and a ISM-like dust-to-gas mass ratio \citep{Klein:2003,Scholz:2006,Harvey:2012,Ricci:2012b}. However, because of the degeneracy between disk radius and disk surface density and geometry, photometry alone does not provide strong constraints on the radial distribution of mass in the disk.   
So far, the only direct observational estimate of the radius of a disk around a single young BD has been obtained by \citet{Luhman:2007b}. They combined Spitzer spectroscopy and high-angular resolution imaging with the Hubble Space Telescope for the nearly edge-on disk around 2MASS J04381+2611, and inferred a disk radius of about $20 - 40$ AU from the optical images of the disk in scattered light. Also, \citet{Ricci:2012b} detected CO molecular gas from the disk surrounding the BD $\rho-$Oph 102 with ALMA at 0.87~mm. They derived a range of possible disk radii: an upper limit of $\approx$ 30 AU from the non resolved dust emission and a lower limit of $\approx$ 15 AU from the CO brightness temperature, assuming that this emission is optically thick.

In this Letter, we present the first spatially resolved observations of dust thermal emission from a young BD disk, 2MASS J044427+2512 (hereafter 2M0444). This object \citep[$\approx 0.05~M_{\odot}$, M7.25-spectral type,][]{Luhman:2004}, located in Taurus ($\sim$ 1 Myr), is known to harbor a disk from its IR excess \citep{Knapp:2004,Guieu:2007,Bouy:2008}. We targeted this BD disk, the brightest in the 1.3 mm survey by \citet[][]{Scholz:2006}, using high-angular resolution ($\approx 0.16''$) interferometric observations with the Combined Array for Research in Millimeter-wave Astronomy (CARMA) at a wavelength of about 1.3 mm.  

Furthermore, Spitzer spectroscopy of this disk provided evidence for $\mu$m-sized dust grains, indicating that the very first steps of grain growth have taken place in this system \citep{Bouy:2008}. We use our observations and others at sub-mm wavelengths to investigate the presence of larger, mm-sized grains in the outer regions of the 2M0444 disk and test the models of dust evolution in protoplanetary disks.

\section{Observations and data reduction}
\label{sec:obs}

2M0444 was observed with CARMA array configurations A, C and D, providing baselines lengths between about 11 and 1678~m. Observations in D-configuration were obtained on October 15 and 18 2011. Receivers were tuned at a local oscillator (LO) frequency of 227.5 GHz, correspondent to a wavelength of about 1.318~mm. 
We setup the CARMA correlator to have a bandwidth of about 6.8 GHz for the continuum, and to observe at the same time the CO ($J = 2-1$) molecular line, but no detection was found for this line. 
We used the same receivers and correlator setup for observations in C-configuration on January 15 2012.
Observations in A-configuration were carried out on December 10 2011. The LO was set at a slightly different frequency, i.e. 224.0 GHz, but considering the spectral index of 2M0444 (see below) the flux of the source is expected to vary by $ < 4\%$ at the two different LO frequencies. This is well below the estimated uncertainty in the absolute flux calibration which is about $10\%$ for CARMA. 

\begin{figure}
\begin{tabular}{cc}
\includegraphics[scale=0.55,trim= 100 130 0 0]{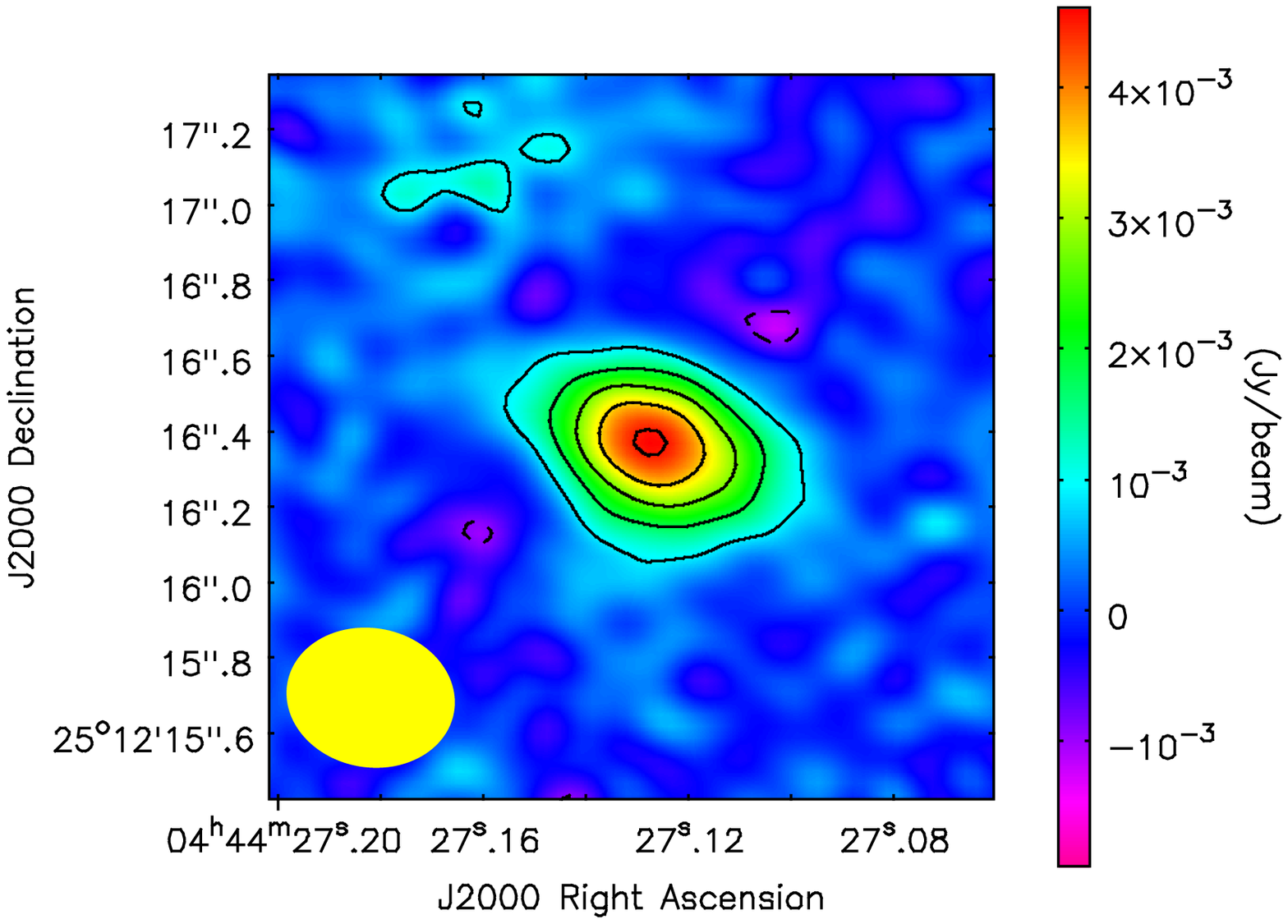} &
\includegraphics[scale=0.55,trim= 50 0 0 0]{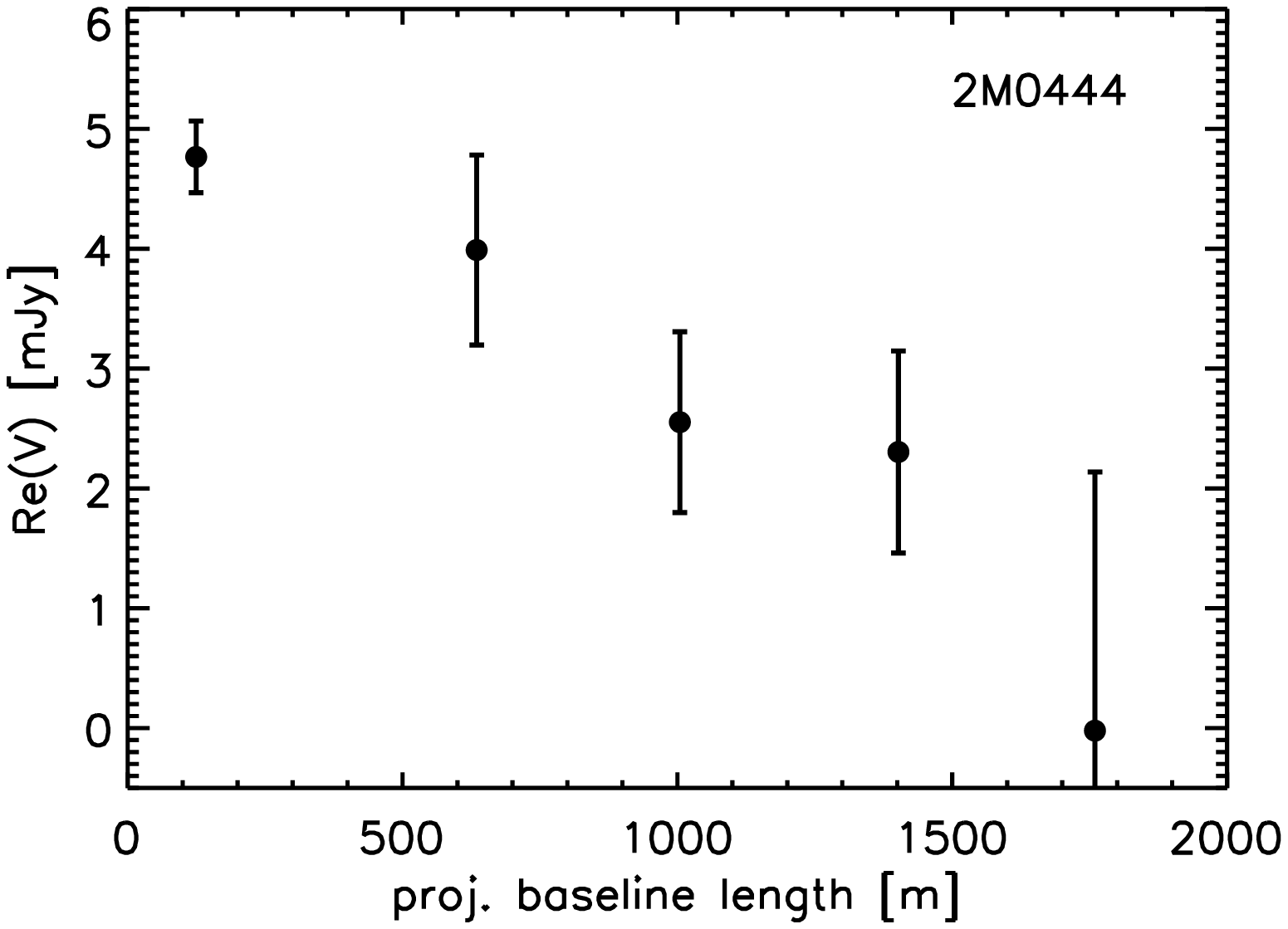} 
\end{tabular}
\caption{Left) CARMA continuum map at 1.3 mm of the disk surrounding the young brown dwarf 2M0444. This map was obtained by combining all the data taken in A, C and D-configuration of CARMA. Contours are drawn at $-3, 3, 6, 9, 12, 15\sigma$, where $1\sigma = 0.3$ mJy/beam is the rms-noise. The yellow ellipse in the bottom left corner shows the synthesized beam, with FWHM $= 0.36'' \times 0.44''$, PA $= 80^{\rm{o}}$, obtained with natural weighting. 
However, the longest projected baselines of our observations probe angular scales as small as 0.16$''$.
Right) Real part of measured visibilities for the 2M0444 disk vs projected baseline lengths. Points and errorbars represent the weighted means of the real parts and their errors, respectively, in the correspondent bins.}
\label{fig:map}
\end{figure}

The frequency-dependent bandpass was calibrated by observing 3C84 and 3C454.3, while the nearby quasar 0510+180 was used to correct for the time-dependent atmospheric and instrumental effects. 
During the observations in A configuration, the quality of these time-dependent corrections were checked by observing the nearby quasar 0431+206, located approximately 5 degrees north of 2M0444. This allowed us to quantify to $\simless~0.05''$ the level of atmospheric seeing during these observations. This is much smaller than the angular resolution of our observations, i.e. $0.16''$, showing that the lower visibilities values at longer baseline lengths (Figure~\ref{fig:map}) cannot be explained by atmospheric decorrelation.  
 
Uranus was used to calibrate the absolute flux scale for observations in D and C configurations, whereas 3C84 was used for data in A configuration, where Uranus is highly resolved. Since 3C84 is known to be a variable source in the millimeter, to evaluate its flux we interpolated between the fluxes measured on December 7 and 15 with the SMA at about 1.3~mm. We obtained a flux density of about 8.3 Jy for 3C84 on the day of our observations.  
For all the four epochs of our observations we shifted the phase center to compensate for the proper motion of 2M0444 \citep[$\mu_{\rm{\alpha}} = - 1.7$ mas/yr, $\mu_{\rm{\delta}} = - 26.5$ mas/yr,][]{Roeser:2010}. 

The MIRIAD software package was used for visibilities calibration and imaging. 
Left panel of Figure~\ref{fig:map} shows the 2M0444 map after combining all the data from different array configurations. 
The integrated flux from the disk is about $5.2 \pm 0.3$ mJy, where the uncertainty reflects the rms noise on the map. 


\section{Analysis of the observed interferometric visibilities}
\label{sec:analysis}

Right panel of Figure~\ref{fig:map} shows the real parts of the measured visibilities vs the projected baseline lengths of our CARMA observations. The decrease at longer baseline lengths indicates that the dust continuum emission of the disk is spatially resolved. 
This sets a lower limit to the angular extent of the disk, namely the angular resolution of the observations. At the Taurus distance of 140~pc, this corresponds to about 22~AU in diameter or 11 AU in radius. 

More precise information on the radial distribution of dust particles can be obtained by analyzing the measured visibilities.
We use the same method adopted by \citet{Isella:2009} to fit interferometric visibilities of proto-planetary disks. They used a Markov Chain Monte Carlo (MCMC) to run large collections of disk models, they calculated for each disk model the predicted visibilities at the ($u,v$)-points sampled by their observations, and then derived estimates for the parameters describing the radial distribution of dust in the disk. 

To analyze the interferometric visibilities of the 2M0444 BD disk, we used two-layer (surface$+$midplane) models of flared disks heated by the radiation of the central BD \citep{Chiang:1997,Dullemond:2001}. For the (sub-)stellar properties of 2M0444 we adopted an effective temperature $T_{\rm{eff}} \approx 2838$ K, a luminosity $L_{\rm{BD}} \approx 0.028~L_{\odot}$ as constrained by \citet{Luhman:2004}, and a distance of 140~pc. By considering the \citet{Baraffe:2003} evolutionary models of BDs and an age of 1~Myr, we derived a mass $M_{\rm{BD}} \approx 50~M_{\rm{Jup}}$. This is within 10\% from the value of $45~M_{\rm{Jup}}$ inferred by \citet{Bouy:2008}. We calculated the dust opacities $\kappa_{\nu}$ by considering porous spherical grains made of astronomical silicates, carbonaceous materials and water ice \citep[optical constants from][respectively]{Weingartner:2001,Zubko:1996,Warren:1984} and used a simplified version of the abundances suggested by \citet[][]{Pollack:1994}, following \citet[][]{Ricci:2010a,Ricci:2010b}. For the grain size distribution we adopted a power-law $n(a) \propto a^{-q}$ with $q=3.0$ and truncated between a minimum size of 0.1 $\mu$m and a maximum size which sets the value of $\beta$~\citep[$\kappa_{\nu} \propto \nu^{\beta}$, e.g.][]{Ricci:2010a}. 

We parametrized the surface density of dust in the disk as a truncated power-law in radius $\Sigma (R) = \Sigma_{\rm{10AU}} (R/\rm{10AU})^{-p}$ from an inner radius $R_{\rm{in}}$ to an outer radius $R_{\rm{out}}$. For the disk inner radius we took $R_{\rm{in}} = 0.05$~AU, which is about the sublimation radius given by the BD properties, but all the parameters considered in our analysis do not significantly depend on the exact value of this parameter. Disk inclination $i$ (angle between disk rotation axis and line-of-sight) and position angle P.A. (angle between north direction and disk major axis) have been left free to vary in the fitting process. However, disk inclination and position angle were not well constrained by our analysis.

For our analysis we first fixed the value of the power-law index $p$ of the radial profile of the dust surface density. We adopted values of $p= 0, 0.5, 1, 1.5$, similar to what found for disks around PMS stars \citep{Guilloteau:2011}. We then launched several MCMC chains to check the convergence toward the absolute minimum of the $\chi^2$-function in the $(\Sigma_{\rm{10AU}}, R_{\rm{out}}, i,$ P.A.$)$ parameter space of the disk models. 
This method allows to efficiently probe the region of parameter space around the $\chi^2$-absolute minimum, and to derive the confidence intervals for the fitting parameters \citep[][]{Isella:2009}.

\begin{figure}
\centering
\begin{tabular}{cc}
\includegraphics[scale=0.5,trim=50 0 0 0]{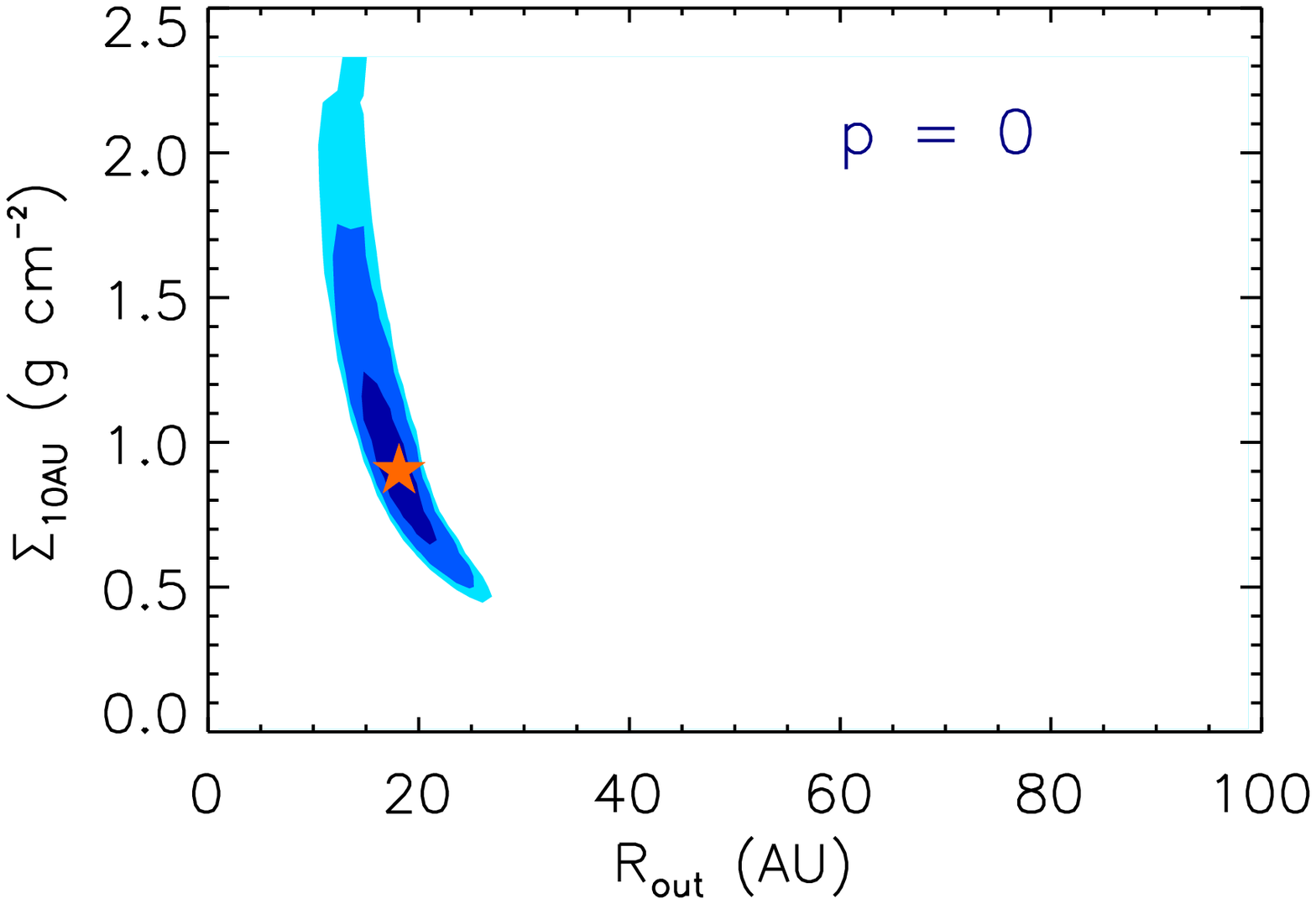} &
\includegraphics[scale=0.5,trim= 0 0 0 0]{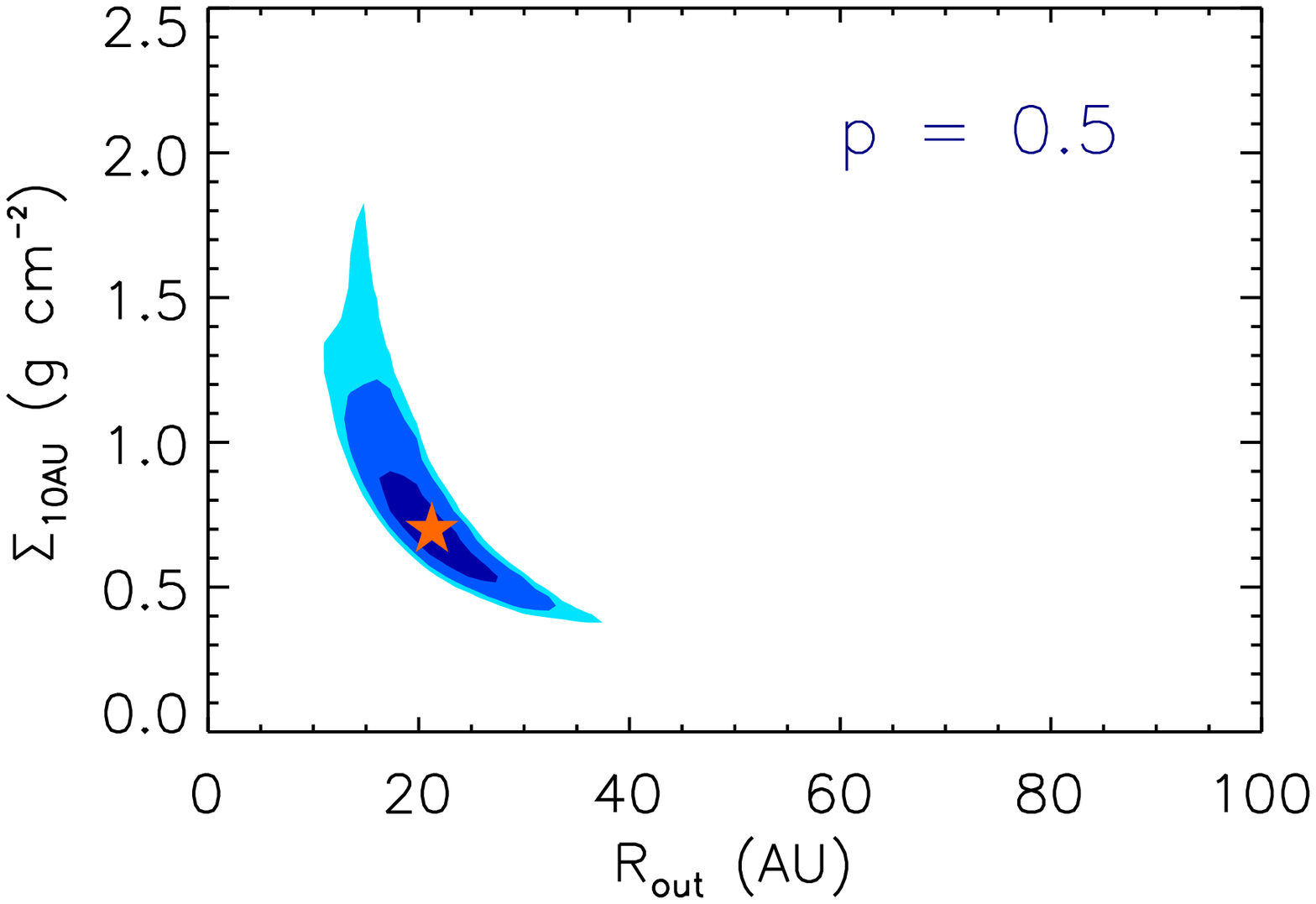} \\
\includegraphics[scale=0.5,trim= 50 0 0 0]{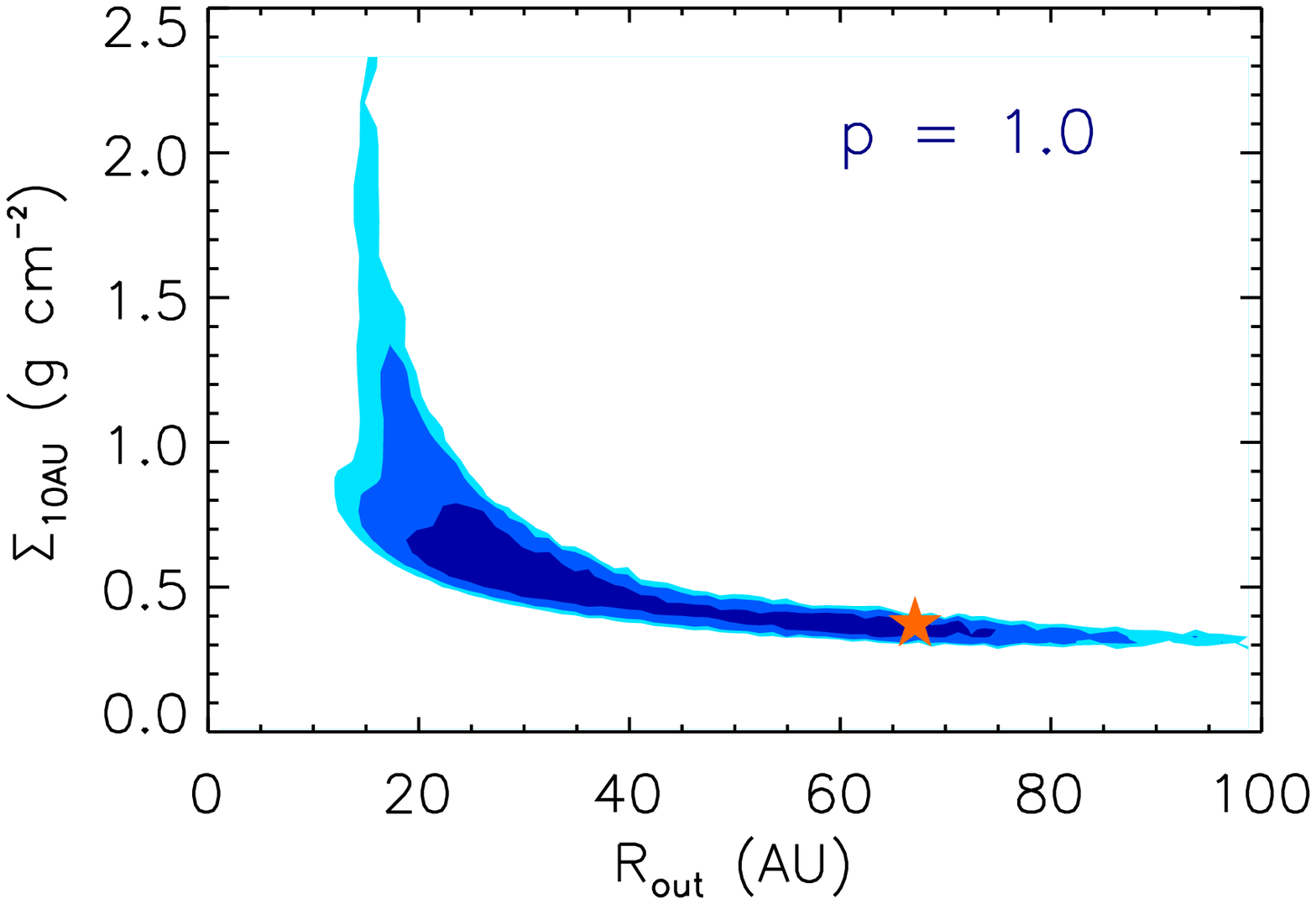} &
\includegraphics[scale=0.5,trim= 0 0 0 0]{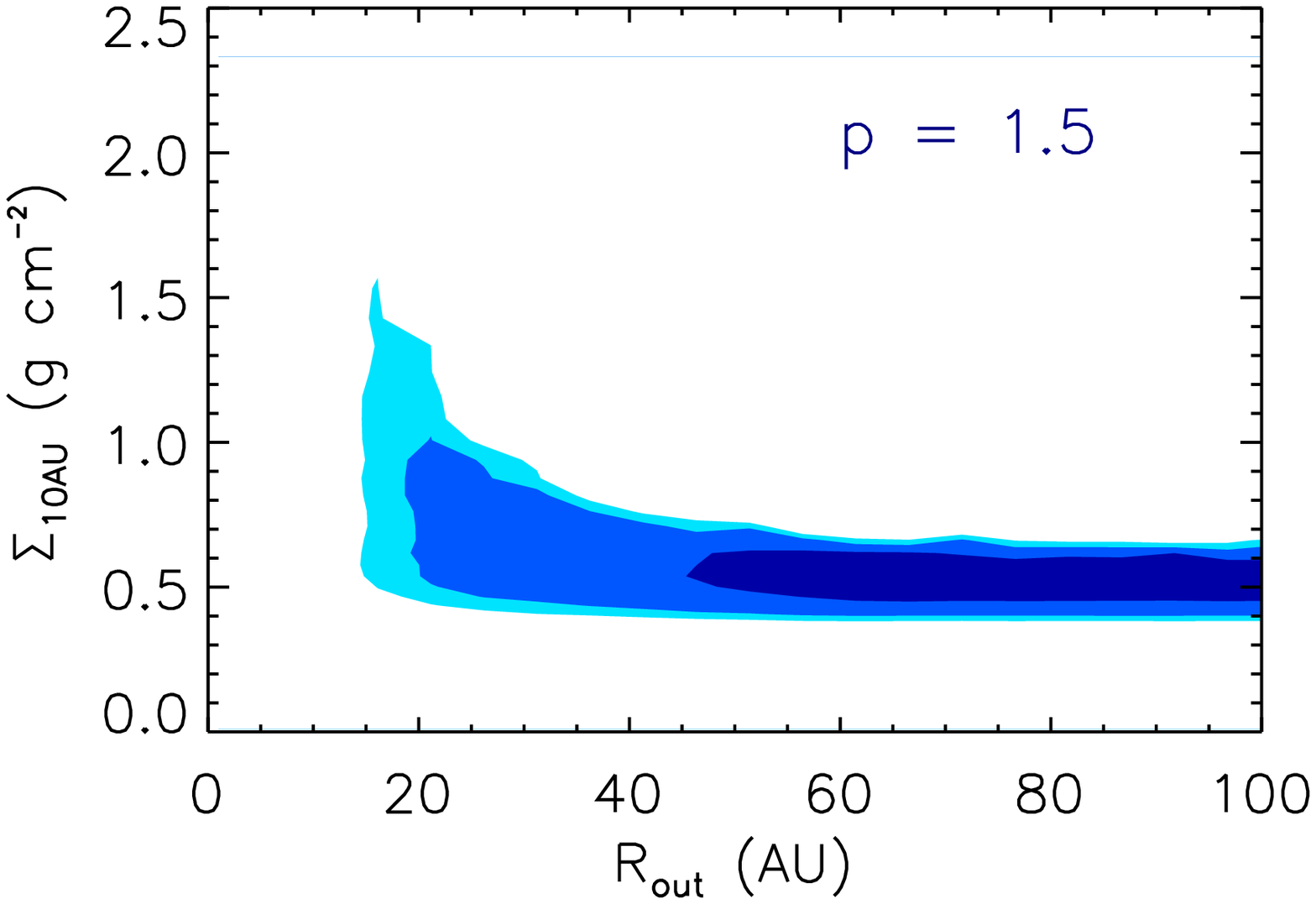} 
\end{tabular}
\caption{Contour plots of the $\chi^2$-function projected into the ($R_{\rm{out}}$, $\Sigma_{\rm{10AU}}$) plane, for different $p$-values. Contours were drawn at $\Delta \chi^2$-values so that their projection on each axis gives the 1D confidence intervals. These are drawn at the $68, 90, 95\%$ confidence levels, going from the darkest to the lightest blue color, respectively. The best-fit models are represented by an orange star.}
\label{fig:contour}
\end{figure}

Figure \ref{fig:contour} shows the 2D contour plots of the $\chi^2$-function on the $(\Sigma_{\rm{10AU}}, R_{\rm{out}})$ parameter space for the different values of $p$. For a given $p$, there is a degeneracy between $\Sigma_{\rm{10AU}}$ and $R_{\rm{out}}$: for larger disks the disk surface density at a given radius has to decrease to conserve the total flux density which is proportional to the disk mass. There is also a clear correlation between $p$ and the disk outer radius of the best-fit models: for disks with steeper radial surface densities, i.e. greater values of $p$, the disk has to be larger to conserve the total flux density, which is dominated by the outermost disk regions. Our observations do not have enough signal-to-noise ratio in the faint outer regions of the disk, and the different $p$-values could not be discriminated by our data. This also means that the confidence intervals on $R_{\rm{out}}$ will depend on the adopted value of $p$. These intervals are listed in Table~\ref{tab:radii}.  
In the case of $p = 0$ or 0.5, the disk outer radius is between about 15 and 27~AU (at $1\sigma$). For larger values of $p$, larger disks are consistent with the data, and in the case of $p = 1.5$ only lower limits for $R_{\rm{out}}$ could be derived.  
The key point we want to highlight here is that for all the reasonable values of $p$ we obtained a robust lower limit for $R_{\rm{out}} > 10$ AU. 

\begin{table*}
\centering \caption{Confidence intervals for the disk outer radii $R_{\rm{out}}$ as constrained by our analysis.} \vskip 0.1cm
\begin{tabular}{ccccc}

\vspace*{1mm} \\

\hline \hline
Confidence level       &      \multicolumn{4}{c}{$R_{\rm{out}}$ (AU)}  \vspace*{1mm} \\
\cline{2-5}
       &     $p=0$        &      $p=0.5$        &       $p=1.0$       &     $p=1.5$          \\
\hline

68\% & 15$-$22 & 17$-$27  &  18$-$74  & $>$ 45   \\
90\% & 12$-$25 & 13$-$33  &  15$-$88  & $>$ 18   \\
95\% & 10$-$27 & 11$-$37  &  12$-$98  & $>$ 15   \\

\hline
\end{tabular}
\begin{flushleft}
\end{flushleft}
\label{tab:radii}
\end{table*}

\subsection{Grain growth in the outer disk}
\label{sec:grain_growth}

Multi-wavelength observations of young disks in the sub-mm have been used to probe mm-sized grains in the disk \citep[][]{Beckwith:1991}. In the case of optically thin emission in the Rayleigh-Jeans part of the spectrum, the flux density of the disk is given by $F_{\nu} \propto \nu^{2+{\beta}}$ where $\beta$ is the power-law index of the dust opacity $\kappa_{\nu} \propto \nu^{\beta}$. Deviations from the Rayleigh-Jeans regime can be accounted for by radiative transfer modeling of the dust emission, while the contribution from optically thick emission can be quantified via resolved images of the disk \citep[][]{Testi:2003}. Whereas tiny sub-micrometer sized grains are characterized by relatively steep sub-mm SEDs with $\beta \approx 1.5 - 2$, shallower spectra with $\beta \approx 0 - 1$ have been typically found for young disks around PMS stars \citep[][]{Andrews:2005,Rodmann:2006,Lommen:2007,Ricci:2010a,Ricci:2010b}. This result has been generally interpreted as evidence for mm-sized grains, or larger, in the outer regions of the disk \citep{Draine:2006,Natta:2007,Ricci:2012a}. 

While grain growth to mm-sized particles have been found for several dozens of disks around PMS stars, the very faint sub-mm fluxes of BD disks make their characterization much more challenging. If compared to typical T Tauri disks, BD disks allow to probe the coagulation of solids to lower gas and dust densities which are the most critical environments for the growth to mm-sized particles \citep[][]{Birnstiel:2010}. Also, models of dust dynamics predict that mm-sized particles should experience a faster inward radial drift in BD disks than in T Tauri disks (Pinilla et al., in prep.). Therefore, BD disks are ideal test beds for models of dust evolution in proto-planetary disks. 

\begin{figure}
 \includegraphics[scale=1]{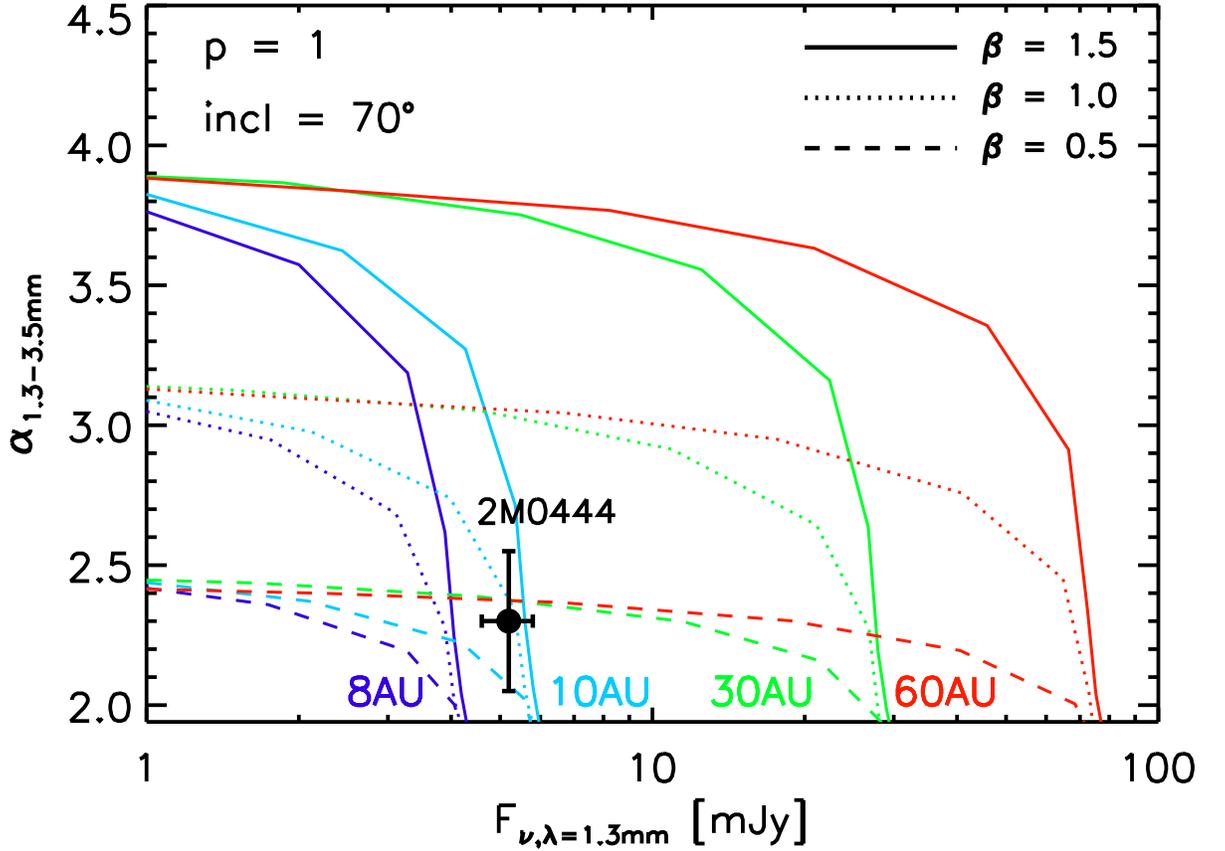}
\caption{Sub-millimeter flux density vs spectral index for disk models for 2M0444. The black dot shows the sub-mm data for 2M0444.
Each line represents the prediction of disk models with the same disk outer radius and dust opacity spectral index $\beta$, as indicated on the plot, but increasing disk mass from left to right. Each model was computed by assuming a radial profile of the surface density $\Sigma \propto r^{-p}$, with $p = 1$, and a disk inclination of 70 degrees. The adopted sub-stellar properties and dust opacities are discussed in Section~\ref{sec:analysis}.}
\label{fig:flux_alpha}
\end{figure}

So far, conclusive evidence for the presence of mm-sized grains in a BD disk has been obtained only for $\rho$ Oph 102 \citep[][]{Ricci:2012b}. They measured a sub-mm spectral index of $2.29 \pm 0.16$, and ruled out small ($< 10$ AU) optically thick disks from the analysis of the molecular CO emission.  
In the case of 2M0444, we measure a sub-mm spectral index of $2.30 \pm 0.25$ by combining our flux density measurement at 1.3 mm with 3.5 mm data obtained by \citet{Bouy:2008}, i.e. $0.55 \pm 0.12$ mJy. Figure~\ref{fig:flux_alpha} shows the measured flux density and sub-mm spectral index overlaid with the predicted sub-mm fluxes for models of disks around a BD with the same properties as 2M0444. The sub-mm photometry of 2M0444 can be explained by two classes of models: relatively small optically thick disks with an outer disk radius $R_{\rm{out}} \approx 10$ AU, and larger, mostly optically thin disks with $R_{\rm{out}} > 10$ AU. In the former case, the disk emission is insensitive to both disk mass and dust opacity $\kappa_{\nu}$, and in fact all the possible values of $\beta$ are consistent with the data.  
This shows how photometry alone is not sufficient to constrain grain sizes in disks~\citep[][]{Testi:2001}. 

At the same time our CARMA observations spatially resolved the dust emission and ruled out models with $R_{\rm{out}} \simless 10$ AU: only the larger, mostly optically thin disk models with $\beta = 0.50 \pm 0.25$ are consistent with both photometric and interferometric data for the 2M0444 BD disk. This interval for $\beta$ is consistent with most of the values constrained for proto-planetary disks around PMS stars, and provides evidence for the presence of particles with sizes $\simgreat~1$ mm in the outer regions of the 2M0444 BD disk. 

Note that Figure~\ref{fig:flux_alpha} refers to disk models with a specific choice of the power law index of the dust surface density and disk inclination, i.e. $p = 1$ and $i = 70$ degrees, respectively. However, these parameters are not observationally constrained. We found that the results presented above are not affected by our ignorance on these parameters by investigating their effect in the ranges of $p \in [0, 1.5]$, $i \in [0, 70]$ degrees\footnote{The upper limit on the disk inclination comes from the low value of visual extinction of 2M0444, i.e. $A_{V} \approx 0$ \citep{Luhman:2004}: a more inclined disk would absorb a fraction of the sub-stellar radiation and would not be consistent with optical spectroscopy~\citep[see][]{Skemer:2011}.}.

\subsection{Disk mass and outer radius: constraints on the mechanisms of BD formation}
\label{sec:disk_radius}

The analysis outlined above shows that the dust emission of the 2M0444 disk is mostly optically thin. This means that a dust mass can be estimated from the sub-mm fluxes. 
The $\beta$-values constrained by our analysis, i.e. $\beta = 0.50 \pm 0.25$, correspond to dust opacities values of $\kappa_{\nu,\lambda=1.3\rm{mm}} \approx 0.2 - 7$~cm$^2/$g$_{\rm{dust}}$ \citep[][]{Ricci:2010a}. We derive estimates for the dust mass of $\sim 1 - 30~M_{\rm{Earth}}$, or $\sim 0.3 - 10~M_{\rm{Jup}}$ in gas 
assuming an ISM-like gas-to-dust mass ratio of 100.
These values correspond to a disk with a mass of $\sim 0.6 - 20\%$ of the mass of the BD, and are consistent with the distribution of values for more massive disks around PMS stars \citep[][]{Williams:2011}.

The constrained values for the disk radius, i.e. $R_{\rm{out}} \approx 15 - 30$ AU for $p =0, 0.5$ or $R_{\rm{out}} > 20$ AU for $p = 1, 1.5$, are in line with previous estimates for BD disks detected
with high-angular resolution optical imaging \citep{Luhman:2007b}, CO molecular line \citep{Ricci:2012b}, sub-mm photometry \citep{Scholz:2006}. These values are very large for models of BD formation involving the ejection of pre-stellar objects in multiple systems, in which dynamical interaction during star-disk and disk-disk encounters significantly modifies the appearance of BD disks \citep[e.g.][]{Reipurth:2001,Bate:2003}. For example, the numerical simulations by \citet{Bate:2003} show that young BDs formed in this way are typically surrounded by relatively small disks with radii $\approx~10$ AU, whereas only $5 - 10\%$ of their BD disks should have $R_{\rm{out}} > 20$ AU. These numbers take in consideration the possible effect of viscous spreading which can lead to an increase of the disk size on a timescale comparable with the disk age. Furthermore, as noted in \citet{Bate:2003}, disks that were severely truncated and underwent following viscous spreading are expected to have a very low mass, because of both truncation and gas accretion during viscous evolution. This is not the case of 2M0444, whose disk-to-central object mass ratio inferred above is similar to what found for disks around PMS stars.   

Here we want to note that the sample of BD disks for which some information on the disk radius has been obtained so far is strongly biased toward the sub-mm bright objects. These might also correspond to the population of relatively large disks predicted by models of BD formation involving ejection processes, whereas the smaller disks with radii $\simless~10$ AU would likely be fainter and therefore not detected at sub-mm wavelengths. A high sensitivity and angular resolution survey of homogeneously selected BD disks in the sub-mm, e.g. with ALMA, is needed to constrain the predictions from these models on a statistical basis \citep[][]{Umbreit:2011} and shed more light on the dominant channel for BD formation.

%
%
%
%

\section{Summary}
\label{sec:summary}

We present high angular resolution observations of the young BD disk 2M0444 obtained with CARMA at 1.3 mm. Our observations spatially resolve dust thermal emission for the first time for a disk around a BD.
By analyzing the interferometric visibilities we infer disk radii of at least a few tens of AU for the 2M0444 disk.
We find evidence for mm-sized grains, or larger, in the disk. Given the physical conditions in BD disks, i.e. lower densities and shorter timescales of solids drift than in typical T Tauri disks, this result provides the tightest test to models of dust evolution in disks, this topic being investigated in detail by Pinilla et al., in preparation.


%
%

\acknowledgments

We thank the OVRO/CARMA staff and CARMA observers for their assistance in obtaining the data. A.I., J.M.C. acknowledge support from NSF grant AST-1109334.
Support for CARMA construction was derived from the Gordon and Betty Moore Foundation, Kenneth T. and Eileen L. Norris Foundation, James S. McDonnell Foundation, Associates of the California Institute of Technology, University of Chicago, states of California, Illinois, Maryland, and NSF. Ongoing CARMA development and operations are supported by NSF under a cooperative agreement, and by the CARMA partner universities. We acknowledge support from the Owens Valley Radio Observatory, which is supported by the NSF grant AST-1140063.

%

\end{document}